\documentclass[lettersize,journal]{IEEEtran}
\usepackage{amsmath,amsfonts}
\usepackage{algorithmic}
\usepackage{array}
\usepackage[caption=false,font=normalsize,labelfont=sf,textfont=sf]{subfig}
\usepackage{textcomp}
\usepackage{stfloats}
\usepackage{url}
\usepackage{verbatim}
\usepackage{graphicx}
\def\BibTeX{{\rm B\kern-.05em{\sc i\kern-.025em b}\kern-.08em
    T\kern-.1667em\lower.7ex\hbox{E}\kern-.125emX}}
\usepackage{balance}
\begin{document}
\title{Explainable AI Framework for COVID-19 Prediction in Different Provinces of India}
\author{Mredulraj S. Pandianchery, Gopalakrishnan E.A, Sowmya V,  Vinayakumar Ravi, Soman K.P
\thanks{Mredulraj S. Pandianchery, Gopalakrishnan E.A, Sowmya V and Soman K.P are with Center for Computational Engineering and Networking (CEN), Amrita School of Engineering, Amrita Vishwa Vidyapeetham, Coimbatore, India. e-mail: mredulraj383@gmail.com, ea\_gopalakrishnan@cb.amrita.edu, kp\_soman@amrita.edu, v\_sowmya@cb.amrita.edu

Vinayakumar Ravi is with Center for Artificial Intelligence, Prince Mohammad Bin Fahd University, Khobar, Saudi Arabia, e-mail: vravi@pmu.edu.sa}}

\maketitle

\begin{abstract}
In 2020, covid-19 virus had reached more than 200 countries. Till December 20th 2021, 221 nations in the world had collectively reported 275M confirmed cases of covid-19 \& total death toll of 5.37M. Many countries which include United States, India, Brazil, United Kingdom, Russia etc were badly affected by covid-19 pandemic due to the large population. The total confirmed cases reported in this country are 51.7M, 34.7M, 22.2M, 11.3M, 10.2M respectively till December 20, 2021. This pandemic can be controlled with the help of precautionary steps by government \& civilians of the country. The early prediction of covid-19 cases helps to track the transmission dynamics \& alert the government to take the necessary precautions. Recurrent Deep learning algorithms is a data driven model which plays a key role to capture the patterns present in time series data. In many literatures, the Recurrent Neural Network (RNN) based model are proposed for the efficient prediction of COVID-19 cases for different provinces. The study in the literature doesn’t involve the interpretation of the model behaviour \& robustness. In this study, The LSTM model is proposed for the efficient prediction of active cases in each provinces of India. The active cases dataset for each province in India is taken from John Hopkin’s publicly available dataset for the duration from 10th June, 2020 to 4th August, 2021. The proposed LSTM model is trained on one state i.e., Maharashtra and tested for rest of the provinces in India. The concept of “Explainable AI” is involved in this study for the better interpretation \& understanding of the model behaviour. The proposed model is used to forecast the active cases in India from 16th December, 2021 to 5th March, 2022. It is notated that there will be a emergence of third wave on January, 2022 in India.
\end{abstract}

\begin{IEEEkeywords}
Covid-19 (novel corona), Time series analysis, prediction, Recurrent Neural Network (RNN), Long Short Term Memory (LSTM), Gated Recurrent Unit (GRU), Explainable AI.
\end{IEEEkeywords}

\section{Introduction}
\IEEEPARstart{I}{n} December 2019, there was an outbreak with unknown cause in the Wuhan city, Hubei province of China. The infection was named as 2019-nCoV on 12th January, 2020 \& the genome sequences were send to World Health Organization (WHO). On 11th March 2020, WHO had declared covid-19 outbreak as global pandemic \cite{1}. 

Artificial intelligence (AI) provides improved methodologies to mitigate the effect of COVID-19 pandemic \cite{13}. The deep learning tools provided by AI are one of the main highlights of improved methodology. Deep learning approaches succeeded to discover the pattern present in large datasets \& had brought breakthrough in visual recognition, speech and video processing \cite{2}. Deep learning approaches plays a prominent role in detecting COVID-19 infections among the infected people \& the early prediction of COVID-19 cases. Deep Learning architectures are used for the detection of COVID-19 infection by the classification of chest radiography images \cite{4}.  They are also used for different real time applications such as face mask detection \& social distance identification in public places which are the most important control measures for the mitigation of COVID-19 \cite{5,6}. Before the implementation of these containment measures in public places, the government forecast the transmission trend of covid-19 among the population. The early prediction of covid-19 plays a key role to control the transmission dynamics of virus among the population. It helps the government to take the necessary precautions \& alert the people by the implementation of different restriction in public places. 

The early prediction of covid-19 cases can be done by the means of mathematical models \& deep learning algorithms. The recurrent deep learning algorithms captures the pattern present in time series data. The forecasting of time series data plays a key role in different domain such as epidemics \& finance \cite{7}-\cite{9}.  Recurrent Neural Network (RNN) based models such as Long Short Term Memory (LSTM) \& Gated Recurrent Unit (GRU) are capable to learn the dynamics in time series data more efficiently as compared to the traditional methods. From the beginning of pandemic, there was lots of study carried related to the transmission trend in different provinces of India \cite{22}. Bahri et. al. proposed LSTM model for the prediction of covid-19 deceased cases in USA, Italy and India. The LSTM model had done the estimation for 7 days ahead with the error rate of 1.37\%, 2.69\% and 1.99\% respectively  \cite{14}. Liu et. al. had proposed the LSTM model for the prediction of cumulative confirmed cases for four different regions in China which include Zhejiang, Guangdong, Beijing, and Shanghai. The proposed LSTM model with LSTM layer of 10 hidden units \& relu as an activation function for the prediction of cumulative confirmed case of next day \cite{17}. Arora et. al. had done the comparison between different deep learning architectures such as Stacked LSTM, Bi-LSTM and Convolutional-LSTM for the daily \& weekly basis prediction of positive COVID-19 cases of each province in India. The study involves training data from 14th March to 8th May, 2020 and testing data from 9th May, 2020 to 14th May 2020. The Bi-LSTM had performed well in daily and weekly prediction among the other deep learning algorithms \cite{21}. ArunKumar et. al. had done the comparison of LSTM \& GRU model with 2 hidden layers \& 300 neuron each in the layer for the prediction \& forecasting of cumulative confirmed cases, cumulative recovered cases \& cumulative fatalities from March, 2020 to October, 2020 in 10 different countries USA, Brazil, India, South Africa, Peru, UK, Brazil, Russia, Mexico, Chile \& Iran separately. The deep learning architectures are trained \& tested on same parent province \cite{27}.

From the literature. It is evident that there are various researches for the efficient prediction of COVID-19 cases. In the literature review, the proposed model is used for the early prediction for different provinces individually i.e., training and testing the model on same parent provinces. The study doesn’t involve whether the proposed model is robust, reliable or interpretable for different stakeholders which include Data Scientist, Product Owners, Regulatory entities, Medical Doctors \& Executive Board Member etc. It is very critical to understand the model behaviour in different tasks such as explanation of predictions in-order to support the decision-making process and debugging the unexpected behaviour of the model. Due to the availability of large databases, well developed methodologies and good computational power, deep learning algorithms perform well on complex tasks. Due to non-linear structure in deep learning models, the decision taken by the neurons are considered to be black box. There is lack of transparency in understanding the model behavior. The requirements of transparency and trust are more crucial in critical applications. Explainable AI is a term which helps to make the results from AI systems more understandable to the stakeholders. Explainable AI contributes to "Responsible AI" considering the accountability, responsibility and transparency \cite{10, 19}. Explainable AI helps to detect the biases existing in the model or data and the weakness existing in the AI system. The detection of improper classification or wrong prediction helps to improve the AI models \cite{11, 12, 18}. The explainable AI approach plays a key role in the medical domain. The explainable AI approach has been used for heart rate variability in ECG signal. The study involves of different deep learning architecture such as LSTM, GRU, CNN and RSCNN trained on one tachycardia disease and tested on different tachycardia diseases. As all tachycardia disease has fast beat rhythm as common characteristic, it will allow to check the robustness \& reliability in the model \cite{20}. 

Motivated by the importance of "Explainable AI", this article is unique from few perspective:
\begin{enumerate}
\item{India is a country with 36 provinces which includes 28 states \& 8 union territories. This study had proposed a LSTM model for active cases per day prediction trained on single state with favourable dynamics \& tested for rest of the provinces. This approach helps to understand the robustness present in behaviour of the model.}
\item{The study also involves the visualization \& interpretation of the biased behaviour detected in model prediction.}
\end{enumerate}

\section{Theoretical Background}

\subsection{COVID-19 Outbreak}
In India, the first case of covid-19 was reported on 27th January, 2020 in Thrissur district of Kerala \cite{28}. During covid-19 outbreak, WHO has acknowledged that the entire world had gazed their attention towards the strategies to be taken by Indian government in-order to control the covid-19 outbreak \cite{29}. The nationwide lockdown was implemented on 24th March, 2020 in India as a precautionary step to control the pandemic \cite{28}. In the absence of vaccine, non pharmacological intervention i.e. social distancing was the only way to control the pandemic. The sudden surge of covid-19 positive cases in hospital has brought the crises in the healthcare operations during the 1st wave of pandemic. The covid-19 outbreak has brought the challenge in the requirements of hospital beds, ventilators, testing services \& front-line workers. It was a great challenge for the country to improve the primary health care system, as 65-68\% of the population lives in the rural areas \cite{30}. The forecasting of covid-19 cases was necessary to capture the transmission dynamics in the different provinces of India. It helps the government to predict the requirements of healthcare facilities \& to make the decision for the implementation of precautionary measures which include lockdown \& quarantine \cite{31}-\cite{33}.

\subsection{Time Series Analysis}
Time Series Analysis is defined as an analysis of orderly sequenced value of the variable which is sampled at the specific interval of time. It captures the evolution of the variable with respect to time \& helps to forecast the future data points \cite{34}. The analysis of time series detect the regularities present in the data which include seasonality \& cyclicity. The different techniques are used for time series forecasting which include regression method, auto regression method, mathematical modelling \& machining learning methods \cite{35}-\cite{37}. The extreme events in dynamical systems are responsible for the sudden transitions in the time series data. This kind of dynamics in the data can be captured with deep learning methods \cite{38}. This study focuses in the implementation of deep learning based model for the prediction of active cases per day in different province of India. 

\subsection{Deep Learning Methods}
Conventional machine learning methods has limitation in its ability to convert the raw data into suitable representation. In the development of machine learning models, domain expertise is needed for the feature extraction from raw data. The representation based learning helps the machine to automatically discover the patterns present in the raw data. Deep learning methods is a representation based learning method which extracts the intricate structure present in the data with multiple levels of abstraction. Deep learning methods had achieved state of art accuracies in different fields which include visual recognition, speech recognition \& natural language processing etc. For time series data, the hand crafted features are more expensive. The recurrent neural networks in deep learning methods helps to capture the past information present in the sequence \& predicts the information in future time steps \cite{2, 39}.

\textit{1) Recurrent Neural Network (RNN):} Recurrent Neural Network captures the pattern present in the data which varies with respect to time. RNN is a structured deep learning model with feedback loop which allows to do the forecasting. In feed forward neural network, there is a unidirectional flow from input layer to the output layer. In RNN, the output from the layer is added with input and fed back into the same layer. RNN takes the sequential input and gives the sequential output from the model. So, RNN plays a key role in sequential modelling with a different variety of applications \cite{24}. The drawback in RNN is that it suffers from the problem of vansishing gradient \& exploding gradient.

\textit{2) Long Short Term Memory (LSTM):} Long Short Term Memory provides gating mechanism which helps to overcome the inefficiency of long term dependencies. LSTM is consist of three important gates which include Input Gate, Forget Gate and Output Gate. The Input Gate in LSTM helps to retain the information in current cell state. The Forget Gate plays a key role in determining which information has to be removed from the previous LSTM cell. The output from LSTM cell is determined by the Output Gate with the combination of cell value, current input and output from the previous iteration \cite{25}. The gating mechanism helps in retaining the relevant information for the next cell and helps to sustain the long term dependency.

\textit{3) Gated Recurrent Unit (GRU):} Gated Recurrent Unit solves the vanishing gradient \& exploding gradient problem in RNN during back-propagation. GRU is consist of two gates which are Update Gate \& Reset Gate. Update Gate determines how much information from the past sequence is relevant for future. Update Gate functions as similar to the Output Gate in LSTM. Reset Gate determines how much information from the past sequence has to be removed. Reset Gate functions as similar to the combination of Input \& Forget Gate in LSTM \cite{40, 41}. GRU does not contain current cell state which is present in LSTM. The information present in current cell state is directly fed in to the hidden state of GRU.

\subsection{Explainable Artificial Intelligence (XAI)}
Now a days, there are lots of development in industrial \& research domain with the help of deep learning. Deep learning had succeeded in solving the critical problems. But the wrong interpretation from the deep learning models is very crucial in different critical domains. So, it is necessary to understand the model behaviour in deep learning approach. The questions always arises in the research with respect to the failure of the model and also how to leverage the performance of the model. The term “Explainable Artificial Intelligence” helps to do the interpretation and understanding from the analysis of the deep learning models. It helps to interpret the decision made in the black box of the neurons. It helps the ML community to have a good interpretability and explainability of ML algorithms \cite{26}. Explainable AI helps to detect the biased decision taken by the deep learning model and also interpret the model behaviour which improve the performance in various field of AI.

\section{Research Methods}
The development of prediction model \& experiments are carried out in Python platform. Different recurrent deep learning architecture are compared for the efficient prediction of covid-19 cases. 

\subsection{Dataset Description}
The dataset for 28 states and 8 union territories of India has been extracted from John Hopkin’s publicly available dataset \cite{23}. The dataset for each province in India is consist of active cases per day, case fatality, cumulative confirmed cases, cumulative death cases, incident rate and cumulative recovered cases. The dataset extracted is for the duration from 10th June, 2020 to 4th August, 2021. Active cases per day is calculated based on cumulative confirmed, deceased and recovered cases. The study involves data which includes 1st wave \& 2nd wave of the COVID-19 pandemic. Many states in India were affected badly during covid-19 pandemic. Maharashtra, Kerala, Karnataka \& Tamil Nadu had reported 20\%, 11\%, 9\% \& 8\% of cumulative confirmed cases in India respectively. Maharashtra, Karnataka, Tamil Nadu \& Delhi had reported 31\%, 9\%, 8\% \& 6\% of cumulative deceased cases in India respectively. Maharashtra is one among the province in India which was badly affected during the pandemic \& also reported maximum active cases per day up to 7,01,614 cases during 2nd wave of pandemic.

\subsection{Model Development}
In this study, deep learning architecture is trained on one province which had reported large number of active cases per day with good dynamics \& this pretrained model is tested for rest of the provinces. This approach helps to understand the robustness in the model and it will also help to improve the results obtained based on the better understanding of the model behaviour. The study also focuses on the comparison between two approaches as follows
\begin{enumerate}
\item{Proposed Approach-1 : To train the model on state with favourable dynamics \& test for rest of the provinces. }
\item{Proposed Approach-2 : To train \& test the model on same parent province.}
\end{enumerate}

\begin{figure*}
 \begin{center}
  \includegraphics[width= 17cm,height=9.5cm]{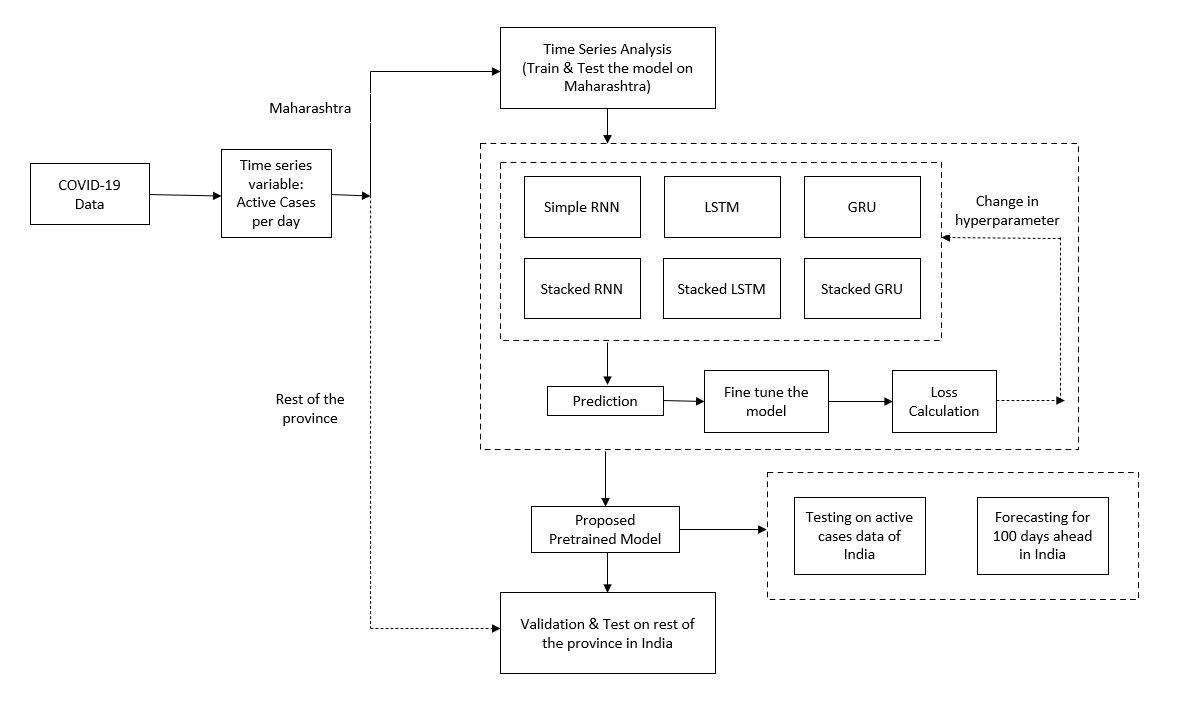}
  \caption{Research Framework.}
 \end{center}
\end{figure*}
As per the data exploration, Maharashtra is one of the states which was badly affected during the 1st and 2nd wave of pandemic in India. Due to the high population density, the covid-19 pandemic had shown the worst effect in Maharashtra. So, the different deep learning architecture is trained on Maharashtra which have favourable dynamics in the data. The different deep learning architecture which include SimpleRNN, LSTM, GRU, Stacked RNN, Stacked LSTM \& Stacked GRU are used for the prediction of active cases per day in Maharashtra. For the further study, the dataset is normalized between 0 \& 1 by MinMax scaler. The normalization is done for data as per the following formula
\begin{equation} \label{eq1}
\begin{split}
{X_{i}} & = \frac{X_{i}-X_{min}}{X_{max}-X_{min}} \\
\end{split}
\end{equation}
where ${X_{max}}$ represents the maximum value of the data \& ${X_{min}}$ is minimum value of the data.
The train dataset is consist of data from 10th June, 2020 to 27th December, 2020 \& the test data from 28th December, 2020 to 4th August, 2021. The deep learning model is trained with 210 days and had tested for 203 days on the dataset of Maharashtra. For Lakshadweep, the 1st case of covid-19 was reported on 18th January, 2021. In proposed approach-2, 300 days are considered in train dataset \& remaining data of Lakshadeep is used to test the model. The deep learning models are trained for 100 epochs with validation split of 10\%. In deep learning model, hard sigmoid function is used as recurrent activation function \& tanh function is used as block input and output activation function. In prediction, loss is measured in MSE (Mean Squared Error) \& optimizer is rmsprop (Root Mean Square Propagation). The metrics for LSTM model is measured in Mean Absolute Error (MAE). MAE can be expressed as 
\begin{equation} \label{eq2}
\begin{split}
MAE & = \frac{\sum_{i=1}^{\infty}{|p_i-a_i|}}{n} \\
\end{split}
\end{equation}
where ${p_i}$ represents the predicted value, ${a_i}$ is actual value \& n is total number of observations. The research framework for training \& testing the model for the efficient prediction of active cases per day is shown in Fig. 1. The pretrained model is also tested on the data of active cases per day for India \& forecasted for 100 days ahead.

\begin{table}
\begin{center}
\caption{Comparison based on MAE (Maharashtra).}
\label{table1}
\begin{tabular}{| c | c | c | c | c | c | c |}
\hline
Sr. & Deep  & No. of &  No. of &  Input & Output & MAE\\
No. & Learning & hidden  & layers & window  & window &  \\
 & Model &  units &  & size & size & \\
\hline
1 & Simple& 150& 1 & 1x7 & 1 & 8866.23\\
 & RNN & &  &  &  &\\
\hline
2 & \textbf{LSTM}& \textbf{150}& \textbf{1} & \textbf{1x8} & \textbf{1} & \textbf{6092.11}\\
\hline
3 & GRU& 100& 1 & 1x8 & 1 & 6494.57\\
\hline
4 & Stacked& 200, & 2 & 1x5 & 1 & 8542.66\\
& RNN & 50 &  &  &  &\\
\hline
5 & Stacked& 150, & 2 & 1x8 & 1 & 7982.91\\
& LSTM & 50 &  &  &  &\\
\hline
6 & Stacked& 50, & 2 & 1x8 & 1 & 7402.05\\
 & GRU & 50 &  &  &  &\\
\hline 
\end{tabular}
\end{center}
\end{table}

\section{Results and Discussion}

\subsection{Selection of Model}

\begin{figure*}[hbt!]
 \begin{center}
  \includegraphics[width= 17cm,height=12cm]{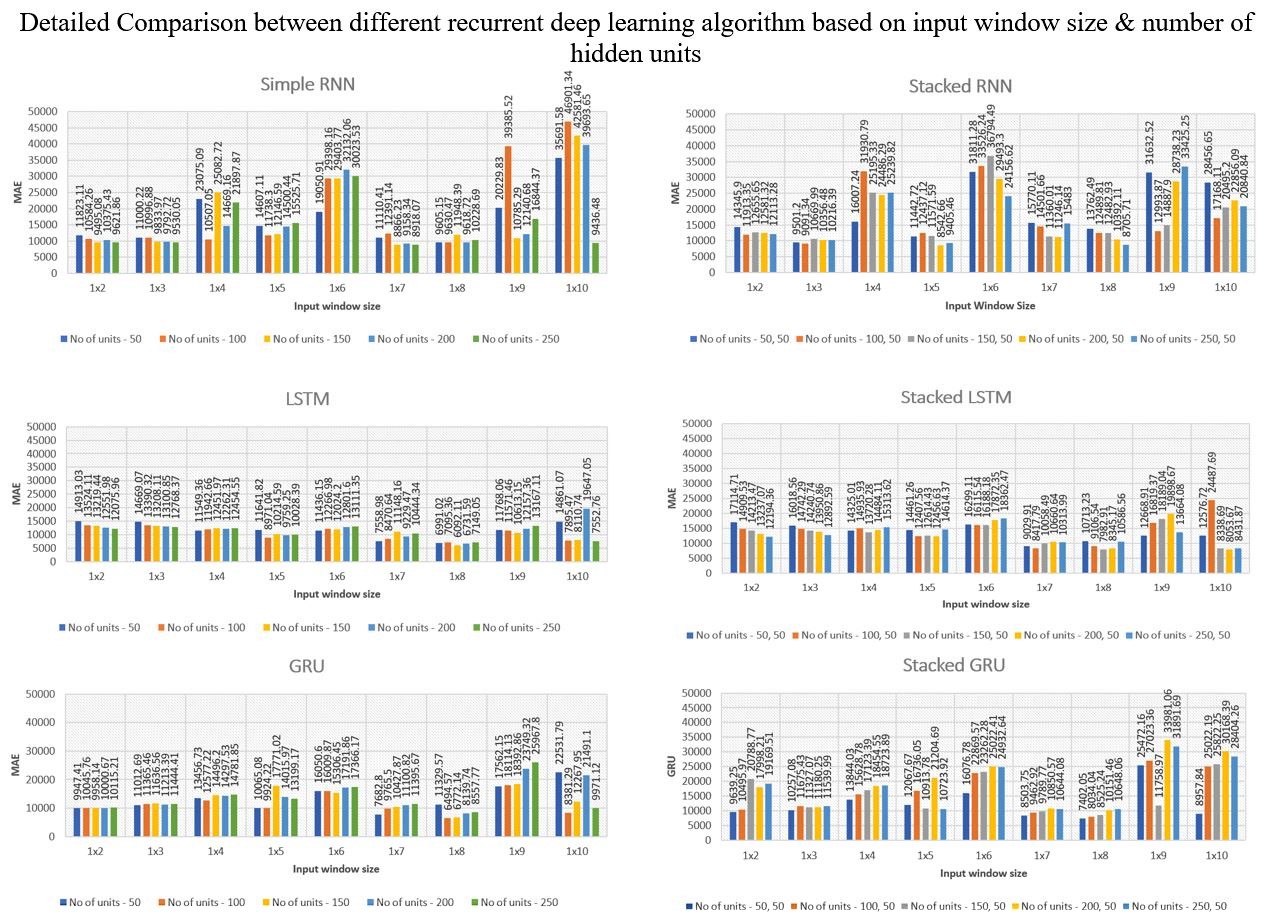}
  \caption{Statistical Heuristics.}
  \includegraphics[width= 18cm,height=6.5cm]{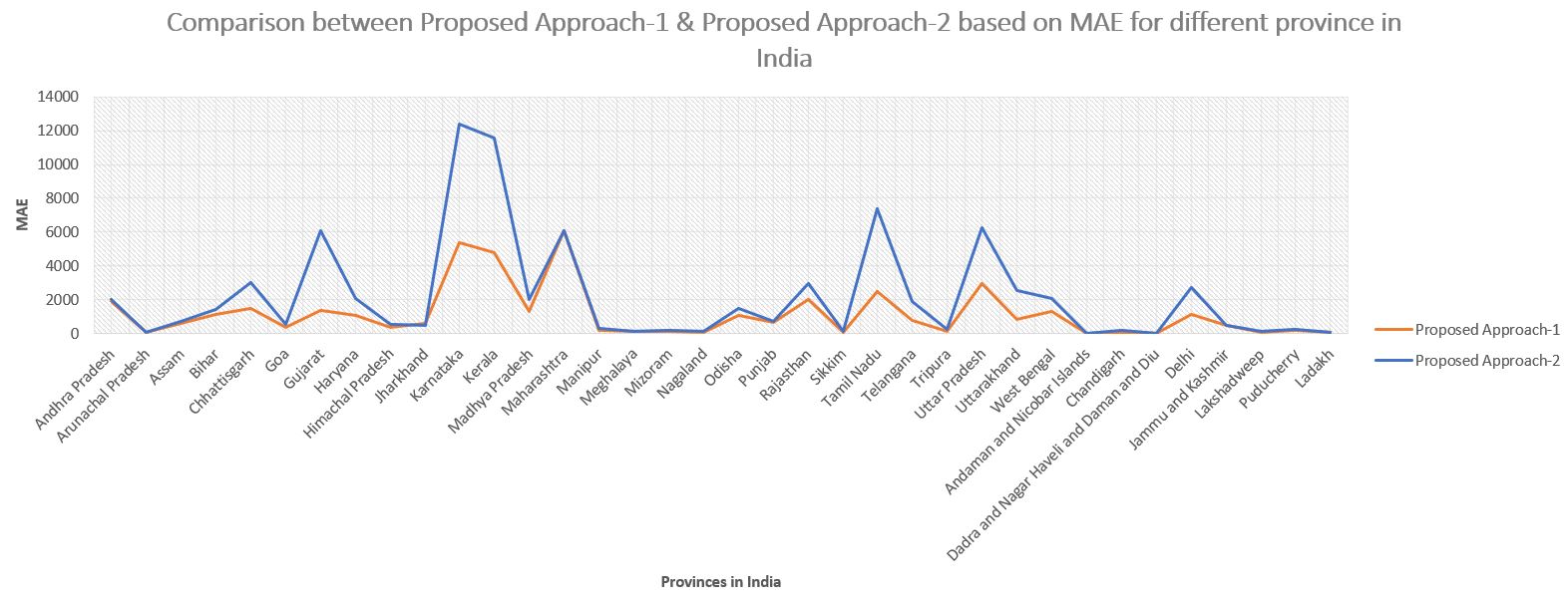}
  \caption{Comparison between Proposed Approach-1 \& 2.}
 \end{center}
\end{figure*}
The comparison is done for different deep learning architecture which include SimpleRNN, LSTM, GRU, Stacked RNN, Stacked LSTM \& Stacked GRU based on MAE obtained after testing the model. The deep learning model with different input size \& number of hidden units are used for the one day prediction ahead. 

Based on comparison with other deep learning architecture in Table \ref{table1}, LSTM model with input size as 1x8 \& number of hidden unit as 150 has given the better result on test data of Maharashtra.  The proposed LSTM model had done the prediction for one day ahead. The statistical heuristics behind the selection of the best model is shown in Fig. 2. The comparison is done between these models based on different input step size \& number of hidden units. The proposed LSTM model which is trained on Maharashtra is used to test for rest of the provinces. The comparison has been done between proposed approach-1 (LSTM model trained on Maharashtra \& tested on rest of the province) \& proposed approach-2 (LSTM model trained and tested on same parent province) based on MAE is shown in Fig. 3. Based on the comparison between MAE of proposed approach-1 \& proposed approach-2, proposed approach-1 had captured the dynamics efficiently as in proposed approach-2 for majority of the provinces in India. 

\subsection{Visualization \& Interpretation of Proposed LSTM Model}

\begin{table}
\begin{center}
\caption{Comparison between proposed approach-1\&2 for 10 provinces based on MAE.}
\label{table2}
\begin{tabular}{| c | c | c | c |}
\hline
Sr. & State  & MAE &  MAE \\
No. &  & (Proposed  & (Proposed \\
 &  &  Approach-1) & Approach-2)  \\
\hline
1& Chhattisgarh & 1496.25 & 2988.53 \\
\hline
2& Gujarat & 1342.59 & 6090.86               \\
\hline
3& Haryana & 1062.18 & 2081.32                \\
\hline
4& Karnataka & 5397.95 & 12373.41           \\
\hline
5& Kerala & 4759.85 & 11581.44          \\
\hline
6& Tamil Nadu & 2504.48 & 7366.15                \\
\hline 
7& Telangana  & 755.34 & 1868.48               \\
\hline 
8& Uttar Pradesh & 2962.28 & 6243.55                 \\
\hline 
9& Uttarakhand & 849.57 & 2539.34          \\
\hline 
10& Delhi & 1135.53 & 2702.94         \\
\hline 
\end{tabular}
\end{center}
\end{table}

\begin{figure}[hbt!]
 \begin{center}
  \includegraphics[width= 6.5cm,height=6cm]{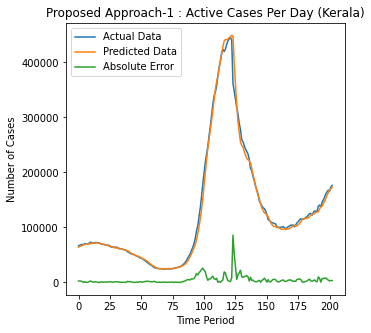}
  \caption{Comparison between Actual \& Predicted data of Kerala in Proposed Approach-1.}
 \end{center}
\end{figure}

\begin{figure}[hbt!]
 \begin{center}
  \includegraphics[width= 7.5cm,height=6cm]{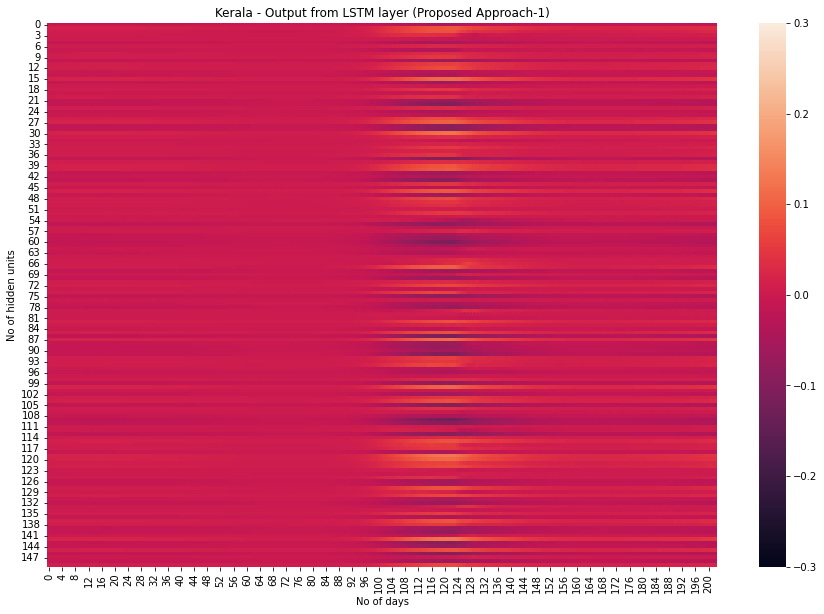}
  \caption{Heatmap for Output from LSTM layer in Proposed Approach-1 (Kerala).}
 \end{center}
\end{figure}

\begin{figure}[hbt!]
 \begin{center}
  \includegraphics[width= 6.5cm,height=6cm]{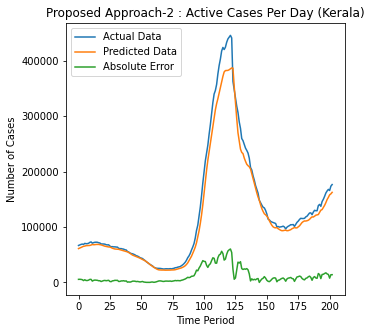}
  \caption{Comparison between Actual \& Predicted data of Kerala in Proposed Approach-2.}
 \end{center}
\end{figure}

\begin{figure}[hbt!]
 \begin{center}
  \includegraphics[width= 7.5cm,height=6cm]{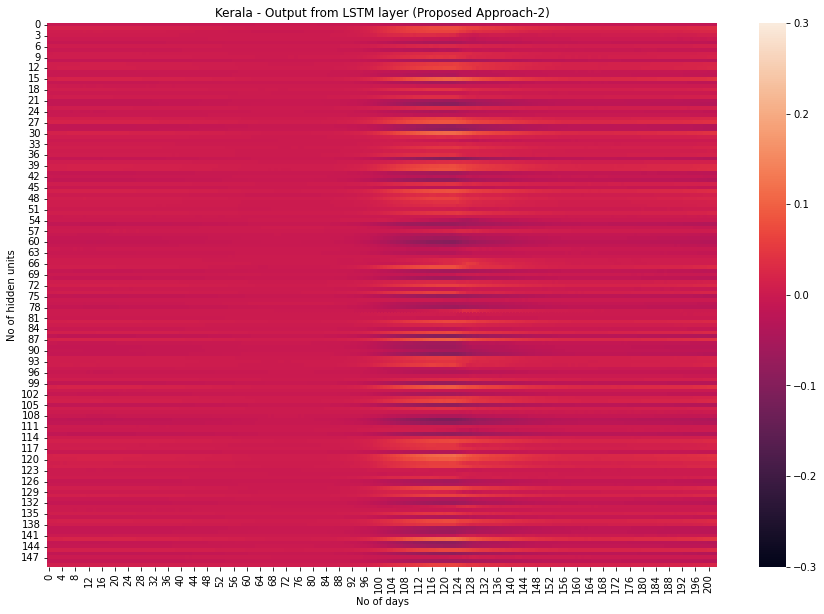}
  \caption{Heatmap for Output from LSTM layer in Proposed Approach-2 (Kerala).}
 \end{center}
\end{figure}

\begin{figure*}[hbt!]
 \begin{center}
  \includegraphics[width= 18cm,height=6cm]{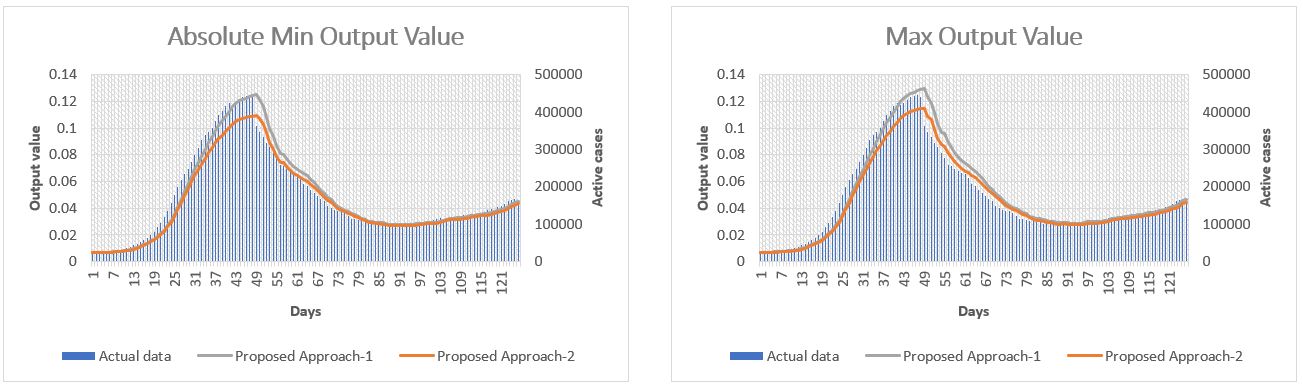}
  \caption{Absolute Minimum Output \& Maximum Output value from LSTM layer (Kerala).}
 \end{center}
\end{figure*}

\begin{figure*}[hbt!]
 \begin{center}
  \includegraphics[width= 14cm,height=6cm]{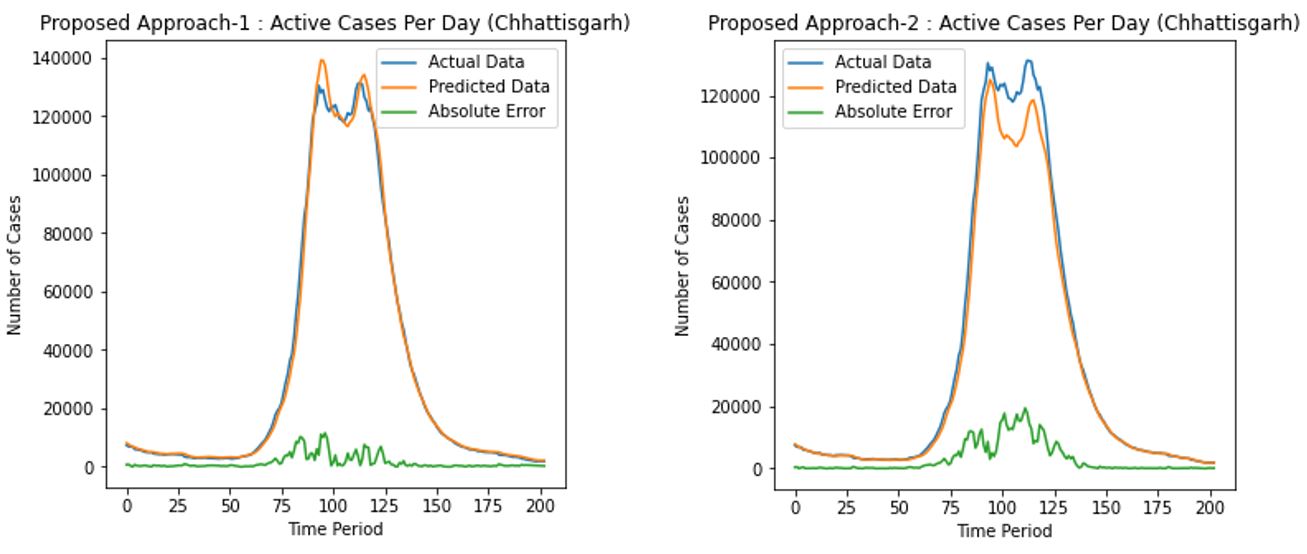}
  \caption{Comparison between Actual \& Predicted data in Proposed Approach-1\&2 (Chhattisgarh).}
 \end{center}
\end{figure*}

\begin{figure*}[hbt!]
 \begin{center}
  \includegraphics[width= 14cm,height=6cm]{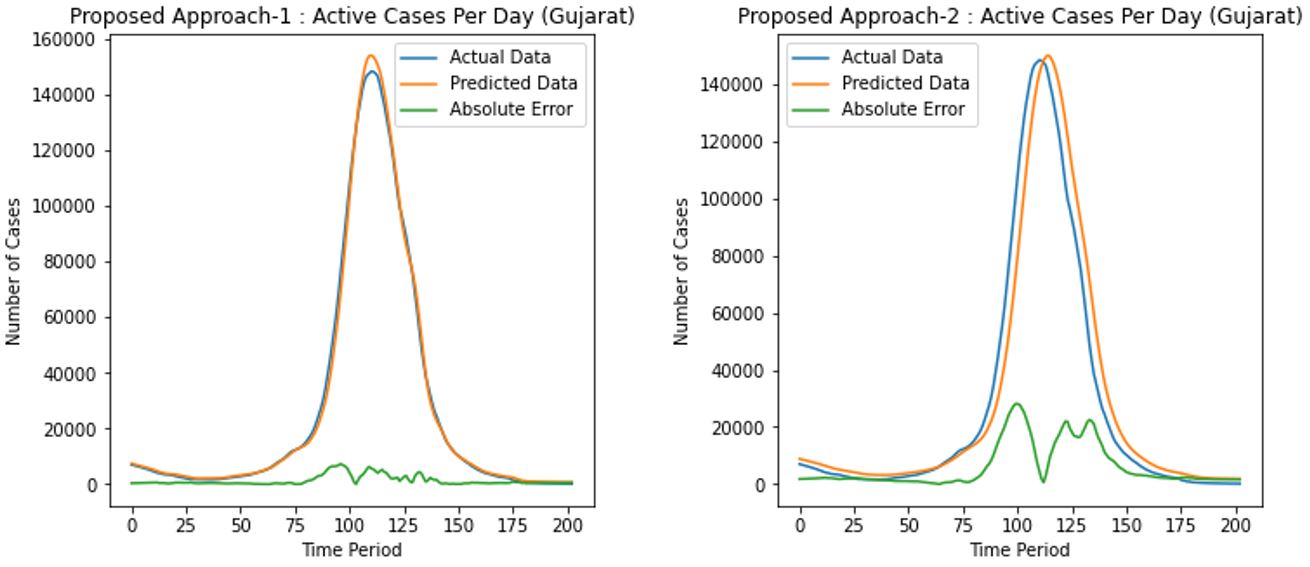}
  \caption{Comparison between Actual \& Predicted data in Proposed Approach-1\&2 (Gujarat)..}
 \end{center}
\end{figure*}

\begin{figure*}[hbt!]
 \begin{center}
  \includegraphics[width= 14cm,height=6cm]{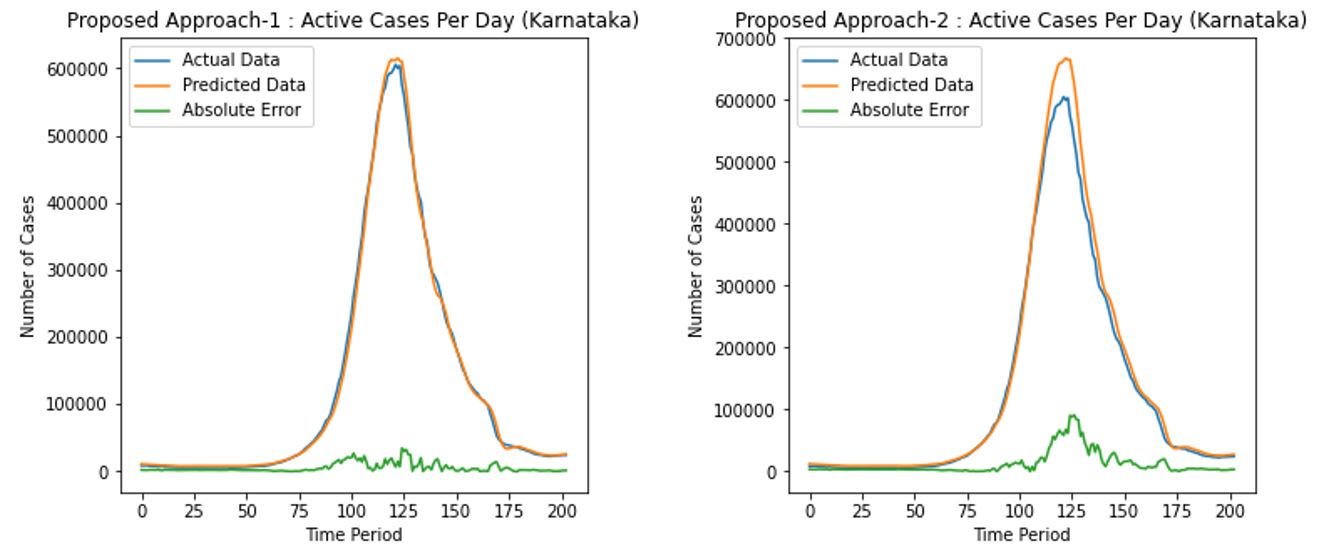}
  \caption{Comparison between Actual \& Predicted data in Proposed Approach-1\&2 (Karnataka)..}
 \end{center}
\end{figure*}

\begin{figure*}[hbt!]
 \begin{center}
  \includegraphics[width= 14cm,height=6cm]{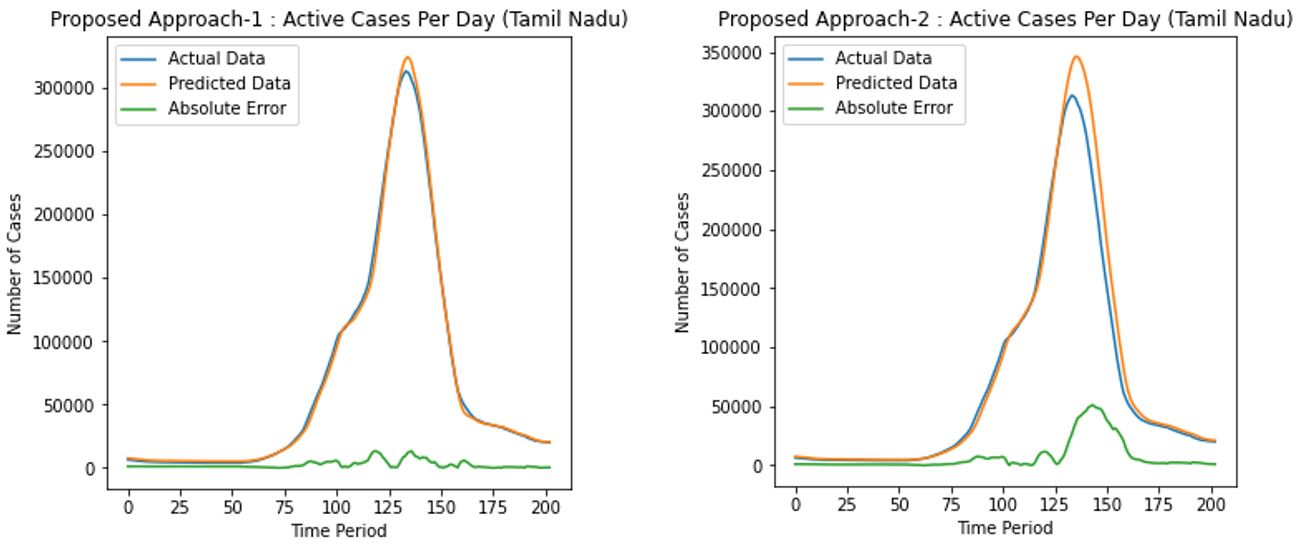}
  \caption{Comparison between Actual \& Predicted data in Proposed Approach-1\&2 (Tamil Nadu)..}
 \end{center}
\end{figure*}

\begin{figure*}[hbt!]
 \begin{center}
  \includegraphics[width= 18cm,height=6cm]{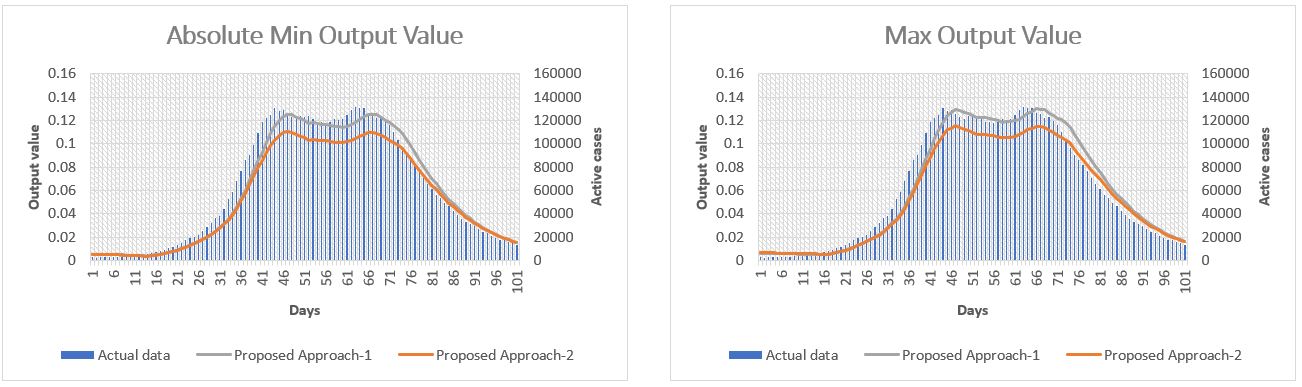}
  \caption{Absolute Minimum Output \& Maximum Output value from LSTM layer (Chhattisgarh).}
 \end{center}
\end{figure*}

\begin{figure*}[hbt!]
 \begin{center}
  \includegraphics[width= 18cm,height=6cm]{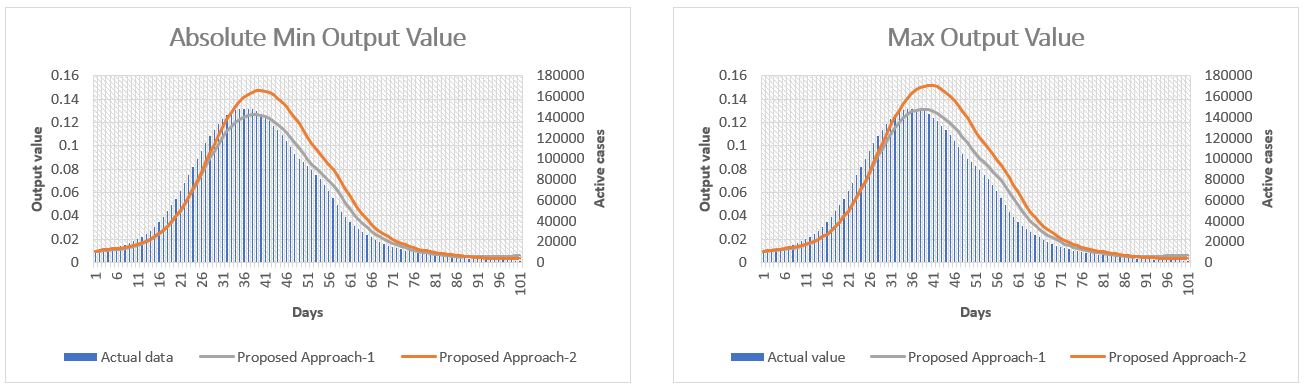}
  \caption{Absolute Minimum Output \& Maximum Output value from LSTM layer (Gujarat).}
  \includegraphics[width= 18cm,height=6cm]{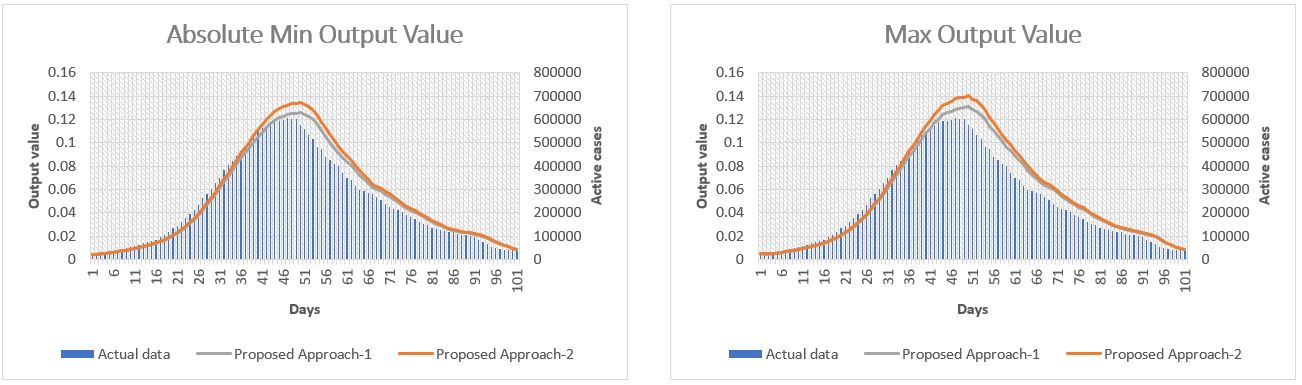}
  \caption{Absolute Minimum Output \& Maximum Output value from LSTM layer (Karnataka).}
  \includegraphics[width= 18cm,height=6cm]{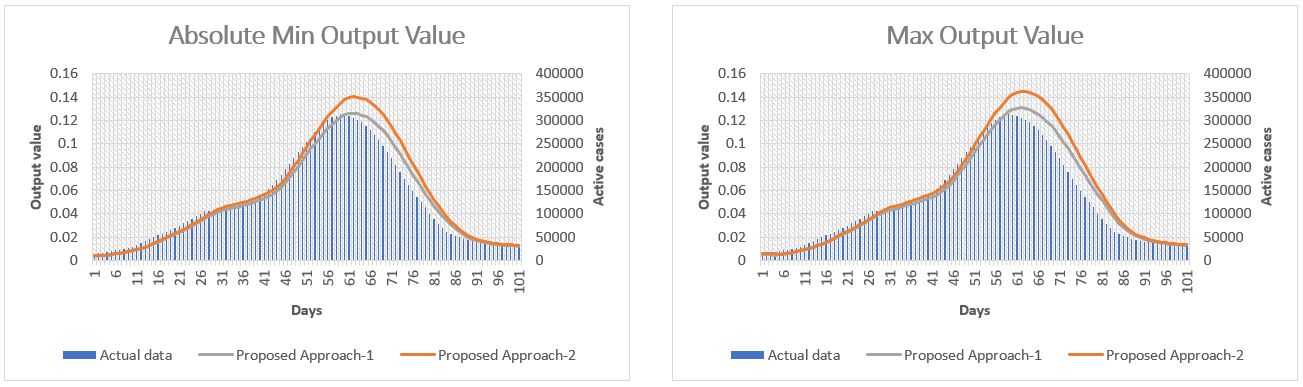}
  \caption{Absolute Minimum Output \& Maximum Output value from LSTM layer (Tamil Nadu).}
 \end{center}
\end{figure*}

From the comparison, it has been notated that proposed approach-1 had performed better as compared to proposed approach-2 for some provinces which include Chhattisgarh, Gujarat, Haryana, Karnataka, Kerala, Tamil Nadu, Telangana, Uttar Pradesh, Uttarakhand \& Delhi with difference in MAE greater than 1000. The MAE calculated for the prediction of active cases per day for these ten states in proposed approach-1\&2 is shown in Table \ref{table2}. In both the approaches, the LSTM model had incorporated one LSTM layer with 150 hidden units, “tanh” as input/output activation function \& “hard sigmoid” as recurrent activation function. So, the study had focussed in the visualization of the output from the LSTM layer. The output from the LSTM layer is elementwise multiplication of output from tanh function (input/output activation function) \& hard sigmoid function (recurrent activation function). At the input of the model, the input step size is 1 x 8 which is mapped to the vector dimension of 1 x 150 at the output of the LSTM layer. For the prediction of 203 days in test data, there will be 203 x 150 dimensional matrix at the output of the LSTM layer. 

The comparison between actual \& predicted data for Kerala in proposed approach-1 \& 2 is shown in Fig. 4 \& 6 respectively. The heat map for 203 x 150 dimensional matrix at the output of the LSTM layer for proposed approach-1 \& 2 are shown in Fig. 5 \& 7 respectively. There is slight variation in the color intensity of heatmap for proposed approach-1 \& 2. The output from the LSTM layer is used to get the prediction at the output of the dense layer in the proposed LSTM model. In proposed approach-2, the large value of absolute error between actual \& predicted data has been notated between 75th and 203rd day of prediction while the model is tested. So, the output value of LSTM layer from 75th to 203rd day of prediction is considered for further study of proposed approach-1\&2. It is difficult to differentiate the heatmaps of proposed approach-1\&2 visually. To differentiate the heatmaps, the maximum \& minimum output value from LSTM layer for each hidden unit is extracted. These two extreme output values have been compared with the actual data to check whether the LSTM layer output had captured the pattern present in the actual data. From the Fig. 8, it is evident that proposed approach-1 had captured the pattern in data in a better way as compared to the approach-2 for Kerala. The pattern present after the peak value in the actual test data has been captured well in maximum activation output as compared to minimum activation output. The pattern present before the peak value in the actual test data has been captured well in minimum  output. The LSTM model in proposed approach-2 had not captured the trend up to the peak value present in actual test data which lead to the large value of MAE between actual \& predicted data from 75th to 203rd day.

In proposed approach-2 for Chhattisgarh, the large value of absolute error between actual \& predicted data has been notated between 50th and 150th day of prediction as shown in Fig. 9. The minimum \& maximum output value from the LSTM layer is shown with actual trend of data in Fig. 13 for Chhattisgarh. So, the output value of LSTM layer from 50th to 150th day of prediction is considered for further study of proposed approach-1\&2. The maximum \& minimum output value had captured the trend in the actual test data in proposed approach-1\&2. While in proposed approach-2, there is a lack of scalability in the two extreme value of  output from LSTM layer as compared to the proposed approach-1 which is also reflected in the predicted value. For states which include Gujarat, Karnataka \& Tamil Nadu. the maximum difference in actual \& predicted data is notated between 75th to 175th day of prediction in proposed approach-2 as shown in Fig 10, 11 \& 12 respectively. From the visualization of maximum \& minimum output of LSTM layer for Gujarat in Fig. 14, it is evident that there is right drift in the pattern captured in proposed approach-2 as compared to approach-1. The right drift in the trend of predicted value is notated for proposed approach-2 in Fig 10. For Karnataka, the trend in actual data is captured in both proposed approach-1 \& 2. There is a slight deviation at the peak value of maximum \& minimum output from LSTM layer in proposed approach-2 as shown in Fig. 15. For Tamil Nadu, there is a deviation in the predicted data after the peak value for proposed approach-2 as shown in Fig. 12. The same deviation of trend has been notated in maximum and minimum activation output in proposed approach-2 with respect to proposed approach-1 and actual trend of test data. The pretrained LSTM model had captured the trend of active cases per day in test data for 36 provinces in India. The visualization of maximum \& minimum output from LSTM layer for proposed approach-1 had shown the robustness in the model to capture the different dynamics. 

\begin{figure*}[hbt!]
 \begin{center}
  \includegraphics[width= 16cm,height=7.5cm]{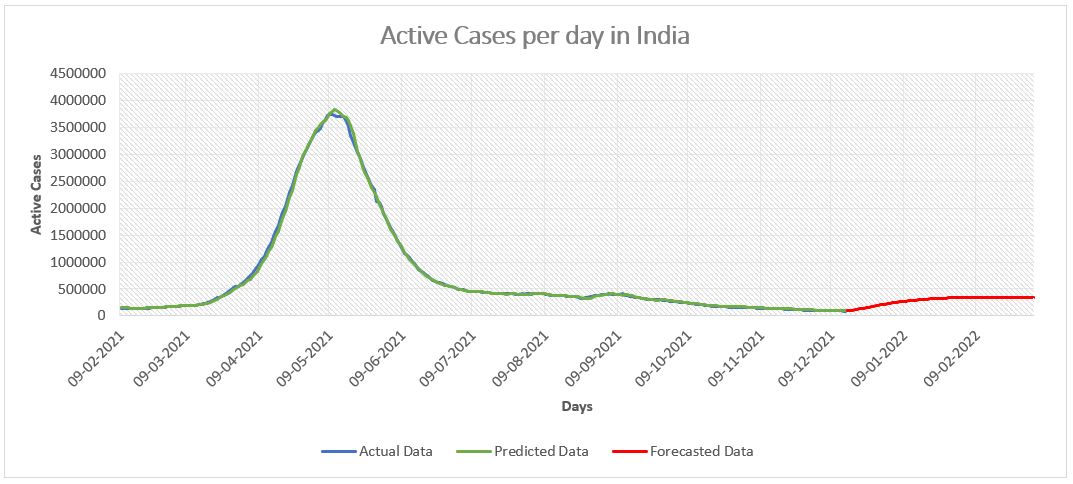}
  \caption{Forecasting of active cases per day in India.}
 \end{center}
\end{figure*}
\subsection{Forecasting of active cases per day for India}

The proposed pretrained LSTM model is used for the forecasting of active cases per day in India. The data of active cases per day for India is collected from https://prsindia.org/covid-19/cases. The data for active cases in India from 9th February, 2021 to 15th December, 2021 is used to test the LSTM model. The MAE calculated between actual \& predicted data for active cases in India is 19547.95. The forecasting of active cases per day is done from 16th December, 2021 to 5th March, 2022 as shown in Fig. 17. As per the forecasting, there will be a rise in the active cases of covid-19 in January \& February 2022. The active cases in India will spike more than 3,00,000 cases. The different variants of SARS-CoV-2 i.e. Beta, Delta \& Omicron is a major threat to entire countries in the world \cite{42,43}. As per the forecasting, there will be a emergence of third wave of pandemic in India from January, 2022. The severity of third wave will be less as compared to second wave of pandemic because of the massive vaccination drive in India. In India, three vaccines which include Covishield, Covaxin \& Sputnik has administered more than 1.37 billion doses (Source: https://www.mohfw.gov.in). More than 0.82 billion \& 0.54 billion population is inoculated with 1st \& 2nd dose respectively. Apart from vaccination, it is necessary to follow the precautionary steps in daily life which include social distancing, use of face mask \& hand-sanitizer to overcome this pandemic \cite{44}.

\subsection{Discussion of Findings}
\begin{enumerate}
\item{The prediction of active cases per day helps the government to be aware about the upcoming wave of covid-19 pandemic. It is necessary to develop the model which can predict the different transmission trend of covid-19 for 36 provinces in India because of distinct geographical location \& population density }
\item{As per the data exploration, Maharashtra is one of the state which was badly affected during 1st \& 2nd wave of pandemic due to the high population density.}
\item{Based on train \& test on the dataset of Maharashtra, the vanilla LSTM had performed better as compared to SimpleRNN, GRU, Stacked version of RNN, LSTM \& GRU.}
\item{The pretrained LSTM model from proposed approach-1 had done the better prediction on test data of all provinces in India. Based on the visualization of maximum \& minimum output from LSTM layer, the pretrained model is robust enough to capture the transmission dynamics of covid-19 in different province of India.}
\item{The forecasting of active cases in India is done by pretrained LSTM model from 16th December, 2020 to 5th March, 2022. The forecast result shows that there will be a spike in active cases by the end of December, 2021.}
\end{enumerate}

\section{Implications of the Study}
\subsection{Theoretical Contributions}
This article is consist of several implications. First, the article had proposed the LSTM model which is trained on single state with favourable dynamics i.e. Maharashtra \& had done the prediction for rest of the provinces. Second, the performance of proposed LSTM model is compared with different model architecture which include SimpleRNN, GRU, Stacked RNN, Stacked LSTM \& Stacked GRU based on different input window size \& number of hidden units. Third, under the importance of Explainable AI, the visualization is done for the maximum \& minimum output value of LSTM layer \& compared with actual trend in the data. Fourth, this article had proved the robustness in the proposed model \& also determined the better results in prediction by proposed approach-1. Fifth, the proposed pretrained model had forecasted the active cases in India for 100 days ahead.

\subsection{Limitations of the Study \& Future Directions}
This article had proposed LSTM model \& new approach for the efficient prediction of active cases per day for different province in India. By the approach of training the model on province with favourable dynamics \& testing on rest of the provinces had helped to enhance the results of prediction. The proposed model had shown the alarm with respect to the spike in active cases of covid-19 in India. It alerts the government to take an appropriate decision \& efficient implementation of precautionary measures. There are many external factors which include mutations of covid-19 virus, human immigration, festival season, political \& religious event,  triggers the transmission of covid-19 among the population. The transmission dynamics of covid-19 changes with the occurrence of such events. So it is necessary to retrain the model as per the change in dynamics. The proposed LSTM model has not incorporated the availability of testing facilities, hospital beds \& ventilators \cite{45, 46}. The health infrastructure plays a key role to control the pandemic. The forecast model incorporated with medical facilities will help to do the efficient prediction of active cases \& requirements in health infrastructure. The another limitation in this article is that the proposed model is not linked with contact tracing of covid-19 infected people. This factors incorporated in the model will helps to efficiently capture the transmission dynamics of covid-19. 

\section{Conclusion}
In this study, the proposed pretrained LSTM model is used for the prediction of active cases in different province of India. The different deep learning architecture which include SimpleRNN, LSTM, GRU, Stacked RNN, Stacked LSTM \& Stacked GRU are trained \& tested on data of Maharashtra. The detailed comparison has been done between these architecture based on different input window size \& number of hidden units. The LSTM model has outperformed the other deep learning architectures on the dataset of Maharashtra. The proposed LSTM model trained on data of Maharashtra had captured the transmission dynamic for rest of the province in India. Under the umbrella of Explainable AI (XAI), this kind of approach tests the robustness present in the model to capture the different dynamics. The proposed approach-1 has given the better results as compared to proposed approach-2. For Lakshadweep, the first case was reported on 18th January, 2021. It was notated that number of training days has to be increased in proposed approach-2 as there is negligible dynamics in data. This kind of problem can be overcomed by proposed approach-1. The study had also focused on the visualization of maximum \& minimum output value from LSTM layer for few states which include Kerala, Chhattisgarh, Gujarat, Karnataka \& Tamil Nadu. From the visualization, it is evident that  the proposed approach-1 had succeeded to generalize the model inorder to capture the different dynamics of transmission. The pretrained LSTM model is used to forecast the active cases in India from 16th December, 2021 to 5th March, 2022. The forecasted data had shown that there will be a emergence of third wave by January, 2022 in India. The forecasting of active cases helps the Indian government to implement the necessary restrictions \& carry out the covid-19 testing in airports, bus \& railway stations.

\end{document}